\documentclass[useAMS]{mn2e}
\usepackage{psfig}

\newif\ifAMStwofonts
\AMStwofontstrue

\def\etal{{et al.}}
\def\xte{{\it RXTE}}
\def\asca{{\it ASCA}}
\def\xmm{{\it XMM-Newton}}
\def\pg{{PG 0804+761}}
\def\bhm{{M$_{\rm BH}$}}
\def\rms{{$\sigma^{2}_{rms}$}}
\def\nb{{$\nu_{bf}$}}
\def\nbh{{$\nu_{hfb}$}}
\def\nbl{{$\nu_{lfb}$}}
\def\psdamp{{PSD$_{\rm amp}$}}

\title[X-ray variability in AGN]
  {The scaling of the X--ray variability with black hole mass in AGN}
\author[I.E. Papadakis]
  {I.E.~Papadakis$^{1,2}$ \\
  $^1$ Physics Department, University of Crete, 71 003, Heraklion, Crete,
  Greece \\
  $^2$ IESL, FORTH-Hellas, 71 110,Heraklion, Crete, Greece}

\def\LaTeX{L\kern-.36em\raise.3ex\hbox{a}\kern-.15em
    T\kern-.1667em\lower.7ex\hbox{E}\kern-.125emX}

\begin{document}

\label{firstpage}

\maketitle

\begin{abstract} 
The relation between the $2-10$ keV, long term, excess variance and
AGN black hole mass is considered in this work. A significant
anti-correlation is found between these two quantities in the sense that
the excess variance decreases with increasing black hole mass. This
anti-correlation is consistent with the hypothesis that the $2-10$ keV
power spectrum in AGN follows a power law of slope $-2$ at high
frequencies. It then flattens to a slope of $-1$ below a break frequency,
\nbh, until a second break frequency, \nbl, below which it flattens to a
slope of zero. The ratio \nbh/\nbl\ is equal to $10-30$, similar to the
ratio of the respective frequencies in Cyg X-1. The power spectrum
amplitude in the (frequency $\times$ power) space does not depend on black
hole mass.  Instead it is roughly equal to $0.02$ in all objects. The high
frequency break decreases with increasing black hole mass according to the
relation \nbh$= 1.5\times 10^{-6}/$(M/$10^{7}$ M$_{\odot}$) Hz, in the
case of ``classical"  Seyfert 1 galaxies. The excess variance of NGC~4051,
a Narrow Line Seyfert 1 object, is larger than what is expected for its
black hole mass and X--ray luminosity. This can be explained if its \nbh\
is 20 times larger than the value expected in the case of a ``classical"
Seyfert 1 with the same black hole mass. Finally, the excess variance vs
X--ray luminosity correlation is a byproduct of the excess variance vs
black hole mass correlation, with AGN accreting at $\sim 0.1-0.15$ the
Eddington limit.  These results are consistent with recent results from
the power spectral analysis of AGN. However, as they are based on data
from a few objects only, further investigation is necessary to confirm
that there is indeed a ``universal" power spectrum shape in AGN (in the
sense that the value of the power spectrum parameters of most AGN will be
distributed around the ``canonical" slope, and amplitude values listed
above). One way to achieve this is to determine the excess variance vs
black hole relation more accurately, using data from many more objects.
This will be possible in the near future, since it is easier to measure
the excess variance of archival light curves than to estimate their power
spectrum. The excess variance vs black hole relation can therefore play an
important role in the study of the X--ray variability scaling with black
hole mass in AGN.
\end{abstract}


%%%%%%%%%%%%%%%%%%%%%%%%%%%%
\section{Introduction}
%%%%%%%%%%%%%%%%%%%%%%%%%%%%

Since the beginning of the active galactic nuclei (AGN) X--ray variability
studies it was noticed that more luminous sources show ``slower"
variations. Barr \& Mushotzky (1986) were the first to show that the
``two-folding" time-scale (i.e.  the time-scale for the emitted flux to
change by a factor of two) was faster in lower luminosity objects.  
Later, Lawrence \& Papadakis (1993) and Green \etal\ (1993), using the
results from the power spectral density function (PSD) analysis of the
{\it EXOSAT} ``long looks", showed that the PSD amplitude at a given
frequency decreases with increasing source luminosity. Nandra \etal\
(1997) and George \etal\ (2000), using the so called ``excess variance"
(\rms, i.e. the variance of a light curve normalised by its mean squared
after correcting for the experimental noise) found an anti-correlation
between excess variance and source luminosity.  Similar results were also
presented by Leighly (1999) and Turner \etal\ (1999), who also found that
a particular group of AGN, the so called ``Narrow Line Seyfert 1" (NLS1)
galaxies, show systematically larger variability amplitude when compared
to the same luminosity ``classical" AGN (i.e. Seyferts with predominantly
broad permitted lines, BLS1s). Further progress in the study of the
longer-term X--ray variability has been afforded by \xte, which has
allowed systematic observations over time-scales longer than before (i.e.
months, years). Markowitz \& Edelson (2001), using 300 day long \xte\
light curves of nine Seyfert galaxies, estimated their excess variance and
found a significant anti-correlation between \rms\ and source luminosity
even on these long time-scales.

Most of the previous studies have examined the dependence of the
``variability amplitude" on the source luminosity.  However, during the
last few years, the mass of the central black hole has been estimated for
many AGN (see e.g. Woo \& Urry 2002 for a recent compilation of AGN black
hole mass estimates). It is therefore possible to investigate the
dependence of the variability amplitude on the black hole mass (\bhm), a
fundamental property of AGN. Recent studies that have followed this
approach, have shown that there exists a significant anti-correlation
between \rms\ and \bhm, using \asca\ light curves of BLS1s and NLS1s (Lu
\& Yu, 2001; Bian \& Zao, 2003).

The primary goal of the present work is to investigate the relation
between excess variance and \bhm\ using the \rms\ data of Markowitz \&
Edelson (2001), together with the recent \rms\ measurement of a
radio-quiet quasar, namely \pg\ (Papadakis, Reig, \& Nandra, 2003).
Markowitz \& Edelson (2001) have already shown that \rms\ is strongly
anti-correlated with the source luminosity. It is important though to see
if there exists a correlation between \rms\ and \bhm\ as well.  In this
way, one can relate the observed X--ray variations to the physical system
itself, since \bhm\ is an important physical parameter of AGN. Based on
the fact that the integral of the PSD is equal to the variance of a light
curve, the \rms--\bhm\ relation is used in order to examine the intrinsic
shape of the PSD in AGN, and how this correlates with \bhm.  Obviously, a
single \rms\ value can correspond to different power spectral forms. As a
result, the \rms--\bhm\ relation cannot provide us with a unique answer as
to what is the intrinsic shape of PSD in AGN, and how this might
correlate/change with \bhm. However, {\it assuming} a certain PSD shape,
the study of the \rms--\bhm\ relation can be a powerful tool in assessing
the reality of the assumed PSD shape.

In this work, the \rms--\bhm\ relation is used in order to examine the
recent results from the power spectrum analysis of \xte, \xmm\ and \asca\
light curves of AGN (e.g. Uttley, McHardy,\& Papadakis, 2002; Markowitz
\etal, 2003). These results have increased significantly our knowledge of
the intrinsic PSD shape of AGN. However, high quality PSD estimation is
currently possible for a few objects only, due to the lack of a large
number of data sets that meet the necessary requirements (i.e. high
signal-to-noise, long, well sampled light curves). The results presented
in this work are also based on the use of a small number of objects, which
have high quality, long term \rms\ measurements. However, although the
number of objects with accurate PSD estimation may not increase
significantly in the near future, the study of the \rms--\bhm\ relation
can be extended soon to incorporate the use of \rms\ measurements from a
large number of objects. For example, the are numerous AGN observations
with \xte\ and \asca, which last typically for $\sim 1$ day (assuming a
$\sim 30-40$ ksec exposure time). Although the PSD estimation is difficult
with these light curves, the measurement of their \rms\ is easier. Taking
also into account the wealth of the new data that are currently provided
by \xmm, and {\it CHANDRA}, it will be possible to define the \rms--\bhm\
relation more accurately in the near future. Consequently, the \rms--\bhm\
relation will probably play an important role in the study of the X--ray
variability scaling with \bhm\ in AGN.

%%%%%%%%%%%%%%%%%%%%%%%%%%%%%%%%  
\section{The dependence of long term variability amplitude on BH mass}   
%%%%%%%%%%%%%%%%%%%%%%%%%%%%%%%%

Markowitz \& Edelson (2001) have calculated the excess variance for nine
Seyfert galaxies using 300 day long, \xte\ (PCA), $2-10$ keV light curves
with a uniform $5$ day sampling. Recently, Papadakis \etal\ (2003)
calculated the excess variance of \pg\ using a year long, \xte\ (PCA),
$2-10$ keV light curve with a $3$ day sampling.  The sampling and duration
of the quasar light curve is similar to the respective properties of the
Seyfert galaxy light curves of Markowitz \& Edelson. Therefore the quasar
\rms\ measurement can be used together with the Seyfert values. The \rms\
values, together with \bhm\ estimates, are listed in Table 1. In most
cases, the \bhm\ estimates are taken from Kaspi \etal\ (2000), and they
correspond to the mean of their ``rms FWHM" and ``mean FWHM" estimates.
All objects with a \rms\ estimate are BLS1s except from NGC~4051 and
MCG~6-30-15, which are classified as NLS1s.

\begin{table}
\caption{The excess variance ($\sigma^{2}_{rms}$), PSD break frequency
(\nbh) and black hole mass (\bhm) estimates of the objects studied in this
work. The objects are ranked by their \bhm. }
 
\label{symbols}
\begin{tabular}{@{}llccc}
\hline
Name & Type &$\sigma^{2}_{rms}$ & \nbh\ & BH$_{\rm Mass}$ \\
 & & $(\times 10^{-2})$ & ($\times 10^{-6}$ Hz) & ($\times 10^{7}$
M$_{\odot})$ \\
\hline
Akn 120	& BLS1 & $3.7\pm 0.7$ & $-$ & $18.6 (1)$  \\
PG0804	& BLS1 & $2.1\pm 0.4$ & $-$ & $17.6 (1)$ \\
3C120   & BLS1 & $3.9\pm 0.8$ & $-$ & $13.5 (2)$ \\
NGC~5548& BLS1 & $5.4\pm 1.1$ & $0.6^{+1.9}_{-0.5} (1)$ & $10.9 (1)$ \\
Fairal 9 (F9)& BLS1 & $4.4\pm 0.9$ & $0.4^{+0.23}_{-0.24} (1)$ & $8.2 (1)$ \\
NGC~5506 & NLS1 & $-$ & $51.2^{+49}_{-50.8} (2)$ & $8.8^{*}$ \\
NGC~3516 & BLS1 & $8.7\pm 1.7$ & $2.0^{+3}_{-1} (1)$ & $1.7 (3)$ \\
NGC~4151 & BLS1 & $10.3\pm 2.1$ & $1.3^{+1.9}_{-1.0} (1)$ & $1.4 (1)$ \\
NGC~3783 & BLS1 & $5.3\pm 1.1$ & $4.0^{+6.0}_{-1.5} (1)$ & $1.0 (1)$ \\
Mrk766 & NLS1 & $-$ & $500^{+300}_{-300} (4)$ &  $0.35 (2)$ \\
MCG-6-30-15 & NLS1 & $8.0\pm 1.6$ & $100^{+100}_{-60} (3)$& $0.1 (4)$ \\
NGC~4051 & NLS1 & $20\pm 4.0$ & $800^{+400}_{-300} (5)$ & $0.05 (5)$ \\
NGC~4395 & BLS1 & $-$ & $320^{+320}_{-160} (6)$ & $0.0066 (6)$ \\
Ark564 & NLS1 & $-$ & $1700^{+600}_{-900} (7)$ & $<0.8 (7)$ \\
\hline
\end{tabular}
Column 2 lists the object type (BLS1 or NLS1). Note that NGC~5506 was
recently classified as NLS1 (Nagar \etal, 2002).  All $\sigma^{2}_{rms}$
estimates in Col.~(3) are taken from Markowitz \& Edelson (2001), except
from \pg\ which is taken from Papadakis \etal\ (2003). The numbers in
parenthesis in Col.~(4) correspond to the references for \nbh\ as follows:  
(1) Markowitz \etal\ (2003), (2) Uttley \etal\ (2002), (3) Vaughan \etal\
(2003), (4) Vaughan \& Fabian (2003), (5) McHardy \etal\ (2003), (6) Shih
\etal\ (2003), (7) Papadakis \etal\ (2002). Note that the \nbh\ values of
NGC~5548 and F9, together with their errors, are taken from Table 5
of Markowitz \etal, while the respective estimate of NGC~3783 was
taken from Section 4.3 of the same work. The numbers in parenthesis in
Col.~(5) correspond to the references for \bhm\ as follows: (1) Kaspi
\etal\ (2000), (2) Woo \& Urry (2002), (3) Onken \etal\ (2003), (4)
Uttley \etal\ (2002), (5) Shemmer \etal\ (2003), (6) Filippenko \& Ho
(2003), (7) Collier \etal\ (2001). The BH mass for NGC~5006 is estimated
as explained in the text. Note that the \bhm\ estimate of MCG~6-30-15 
is rather uncertain (see discussion in Section 6.3 of Uttley \etal, 
2002.)
\end{table}

In order to compute the uncertainty of the \rms\ estimates, the
uncertainty of the \pg\ \rms\ measurement was estimated first, based on
the PSD results of Papadakis \etal\ (2003). As \rms\ is roughly equal to
the PSD integral from the highest to the lowest sampled frequency, using
the best-fitting parameter values from the power-law model fitting to the
PSD, it is straightforward to compute the uncertainty of the PSD integral
over the sampled frequencies, and hence \rms\ itself. The result shows
that [error(\rms)/\rms]$\sim 0.2$. The uncertainty of the PSD integral
depends mainly on the frequency range and frequency resolution of the PSD.
These properties are determined by the number of points and the length of
the light curve. Since all the \rms\ values listed in Table 1 are
estimated by light curves with almost identical length and number of
points, the ``signal-to-noise" ratio of the \rms\ values should be $\sim
0.2$ in all cases. The \rms\ errors listed in Table~1 are computed based
on this assumption. As for the \bhm\ estimates, reverberation mapping and
stellar velocity dispersion methods give reliable estimates within factors
of a few (Woo \& Urry, 2002). However, it is difficult to determine the
actual error of all the individual \bhm\ estimates listed in Table~1. For
that reason they will not be considered hereafter.

%%%%%%%%%%%%%% Fig. 1 %%%%%%%%%%%%%%%%%%
\begin{figure}
\psfig{figure=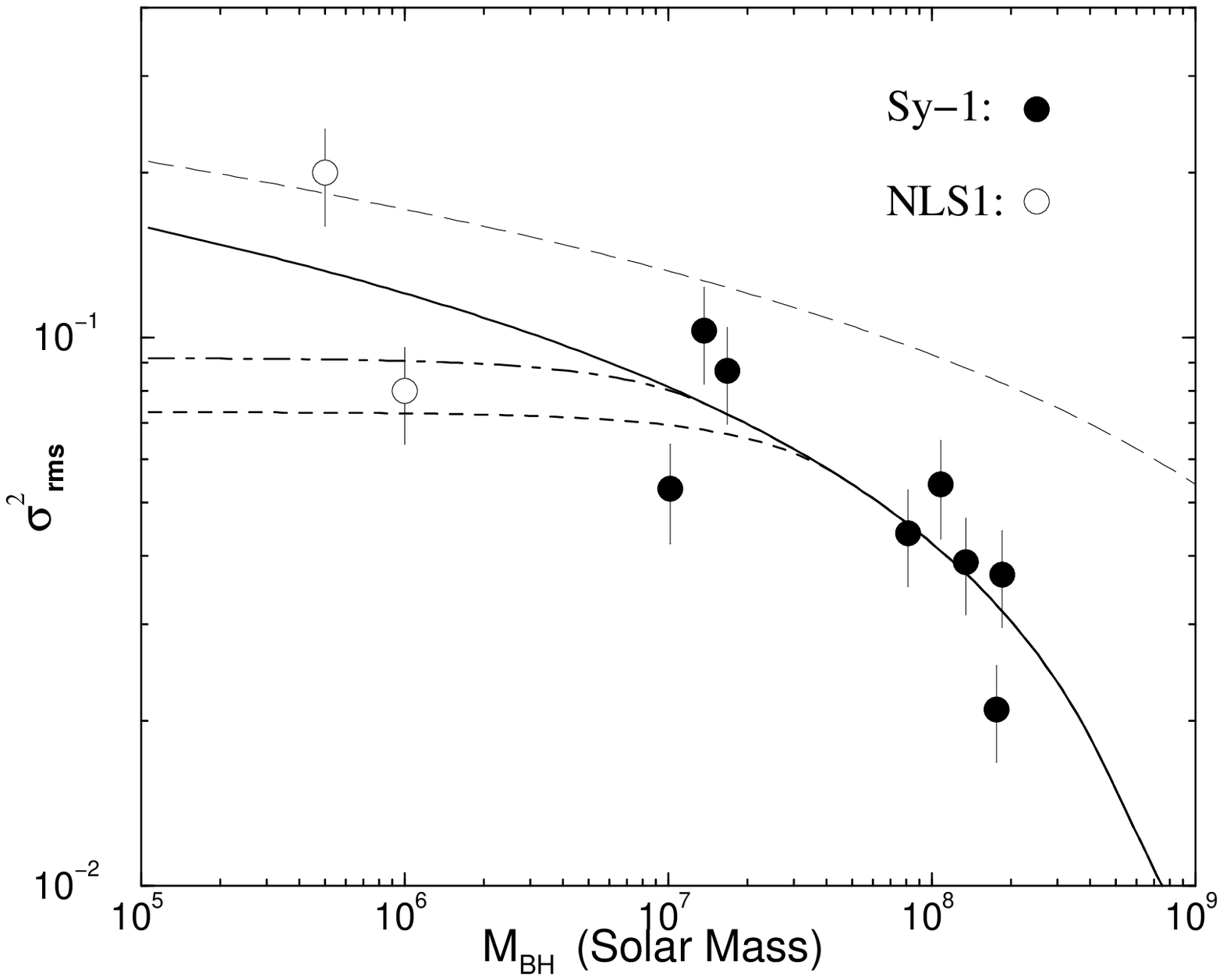,height=7.0truecm,width=9.5truecm,angle=0,%
bbllx=20pt,bblly=10pt,bburx=570pt,bbury=460pt}
\caption[]{\xte\ excess variance plotted as a function of \bhm. The solid
line shows the best-fitting model in the case of a universal PSD with
\nbh\ $\propto $ \bhm$^{-1}$. The long-dashed line shows the \rms\ vs
\bhm\ relation in the case when \nbh\ scales with \bhm\ with a
normalisation 20 times larger than case of the model shown with the
solid line.  The short-dashed and dot-dashed lines show a model with two
PSD break frequencies, \nbh\ and \nbl, with \nbl\ being $10-30$ time 
smaller than \nbh, respectively (see text for details).}
\end{figure}
%%%%%%%%%%%%%%%%%%%%%%%%%%%%%%%%%%%%%%%%

Fig.~1 shows the \rms\ vs \bhm\ plot.  Clearly, there is a strong
correlation between the two variables in the sense that that \rms\
decreases with increasing \bhm. Computation of Kendal's $\tau$ yields
$-0.69$, which implies that the anti-correlation between \rms\ and \bhm\
is highly significant (probability $>99.5\%$). 

Let us suppose that the power spectrum, normalised to the mean squared,
follows the form: $P(\nu)=A(\nu/\nu_{hfb})^{-1}$ Hz$^{-1}$, for
$\nu<\nu_{hfb}$, and the form:  $P(\nu)=A(\nu/\nu_{hfb})^{-2}$ Hz$^{-1}$,
at higher frequencies ($A$ is the PSD value at $\nu=\nu_{hfb}$). Let us
also suppose that the PSD ``amplitude", i.e. \psdamp=$A\times\nu_{hfb}$,
is constant and does not depend on \bhm, while \nbh\ is inversely
proportional to the black hole mass, i.e. \nbh$=C/$M$_{7}$ Hz, where
M$_{7}$=\bhm/$10^{7}$M$_{\odot}$.  Under these assumptions, one can find a
unique relation between \rms\ and M$_{7}$ using the fact that
$\sigma^{2}_{rms} = \int_{\nu_{ls}}^{\infty} P(\nu)d\nu$, where $\nu_{ls}$
is the lowest frequency sampled ($=1/300$ days = $3.858\times10^{-8}$ Hz,
for the objects considered in this work). Note that, strictly speaking,
the upper limit of the \rms\ integral should be closer to $2\Delta
T_{obs}$, where $\Delta T_{obs}\sim 1$ ksec is the length of the
individual monitoring exposures, since variations on shorter time scales
are smoothed out. In any case though, use of either limits does not affect
the integral's value almost at all.

The integral's estimation depends on whether \nbh\ is higher or lower than
$\nu_{ls}$. In the first case, one can show that:

\begin{equation}
\sigma^{2}_{rms} = {\rm PSD_{amp}}\times [\ln (C/{\rm M}_{7}) +18.07],
\end{equation}

\noindent
while in the second case, \rms\ is given by the relation,

\begin{equation}
\sigma^{2}_{rms} = {\rm PSD_{amp}}\times 2.592\times 10^{7}\times 
C/{\rm M}_{7}.
\end{equation}

The data shown in Fig.~1 were fitted with the model given by equations (1)
and (2), excluding the points which correspond to the two NLS1s. The best
fitting model is also shown in Fig~1 (solid line). The best fitting model
parameter values are \psdamp$=0.017\pm0.006$, and
$C=1.7^{+3.9}_{-0.8}\times 10^{-6}$ Hz (errors correspond to the $68\%$
confidence for two interesting parameters, i.e. $\Delta \chi^{2}=\chi^{2}
+ 2.3$). Although the model does not provide a statistically accepted fit
to the data ($\chi^{2}=14.8/6$ dof), it describes rather well the overall
trend of decreasing \rms\ with increasing \bhm. The \rms\ values of the
two NLS1s (NGC~4051 and MCG~6-30-15), NGC~3783, and \pg\ show the largest
deviations from the best fitting model.

%%%%%%%%%%%%%%%%%%%%%%%%%%%%%%
\subsection{The possibility of a second frequency break in the PSDs}
%%%%%%%%%%%%%%%%%%%%%%%%%%%%%%

If there is a second PSD flattening to zero slope at a frequency
\nbl$<$\nbh\ (like the PSD of Galactic black hole candidates, GBHs, when
at low/hard state), then \rms\ is given by the following relation,

\begin{equation}
\sigma^{2}_{rms} = {\rm PSD_{amp}}\times
[2-\nu_{ls}/\nu_{lfb}+\ln(\nu_{hfb})-\ln(\nu_{lfb})],
\end{equation}

\noindent in the case when \nbl$>\nu_{ls}$. In the opposite case, \rms\ is
not affected by the presence of \nbl\, and equations (1), (2) still hold.
The dashed and dot-dashed lines in Fig.~1 show a plot of equation (3)
using the best fitting \psdamp\ and $C$ values from the previous section,
and assuming that \nbh/\nbl$=10-30$, respectively. This ratio is similar
to the ratio of the respective break frequencies in the PSD of Cyg X-1 at
low/hard state (Belloni \& Hasinger, 1990). Equation (3) provides an
acceptable fit to all data (except from NGC~4051): $\chi^{2}=10.7/7$ dof
and $\chi^{2}=10.2/7$ in the case when \nbh/\nbl$=10-30$, respectively.

As Fig.~1 shows, the MCG~6-30-15 and NGC~3783 \rms\ measurements are now
consistent with the model defined by equation (3). In other words, their
\rms\ measurements suggest that there are two breaks in their PSD, both of
which are higher than $\nu_{ls}$. Interestingly, the observed PSD of
NGC~3783 does show two breaks, with \nbh/\nbl$\sim 20$ (Markowitz \etal,
2003). However, only one break (which corresponds to the ``$-2$ to $-1$"
slope flattening) has been reported in the case of MCG~6-30-15. Uttley
\etal\ (2002) and Vaughan, Fabian, \& Nandra (2003), report \nbh$\sim
5\times 10^{-5}$ Hz, and $\sim 1\times 10^{-4}$ Hz, respectively, with the
two values being consistent within the errors. If indeed \nbh\ $\sim
5\times 10^{-5}-10^{-4}$ Hz, the present results suggest that the low
frequency break should be located at a frequency $10-30$ times lower, i.e.
at around $\sim 1-5\times 10^{-6}$ Hz. Such a break should be detected by
Uttley \etal\ (2002), as their PSD extends to frequencies lower than $\sim
10^{-7}$ Hz. A combined long-term \xte\ and short-term \xmm\ PSD could
resolve this issue.

%%%%%%%%%%%%%%%%%%%%%%%%%%%%%%
\subsection{The case of NGC~4051}
%%%%%%%%%%%%%%%%%%%%%%%%%%%%%%

NGC~4051 is the only source which deviates significantly from all model
lines plotted in Fig.~1. This NLS1 is one of the most variable radio-quiet
AGN. The discrepancy between the NGC~4051 \rms\ estimate and the model
defined by equation (3) (dashed and dot-dashed lines in Fig.~1) suggests
that the PSD of this object should have just one frequency break at
frequencies higher than $1/300$ days$^{-1}$. This is consistent with the
results of McHardy \etal\ (2003). Using \xte\ and \xmm\ data, McHardy
\etal\ detect a frequency break at $\sim 8\times 10^{-4}$ Hz (where the
PSD changes slope from $-2$ to $-1$), but they do not observe any further
flattening at lower frequencies down to $10^{-8}$ Hz. However, the
NGC~4051 \rms\ measurement does not agree with the best-fitting model
defined by equations (1) and (2) (solid line in Fig.~1). This implies that
either the \psdamp of NGC~4051 is larger than in other AGN, or \nbh\ is
higher than expected for its \bhm. According to the results of McHardy
\etal\ (2003), \psdamp$\sim 0.01$, which is smaller than the best fitting
\psdamp\ value of $0.017$. Therefore, the \rms\ measurement of NGC~4051
can be consistent with equations (1) and (2) only if $C$ is increased. The
long-dashed line in Fig.~1 shows a plot of the model defined by equations
(1) and (2), using \psdamp$=0.017$, and $C=20$ times larger than the
best-fitting value reported in Section 2. The agreement now between
the model and the NGC~4051 measurement is good.

%%%%%%%%%%%%%%%%%%%%%%%%%%%%%%%%%%%%%
\subsection{The scaling of \nbh\ with \bhm}
%%%%%%%%%%%%%%%%%%%%%%%%%%%%%%%%%%%%%

One of the results from the model fitting of the \rms--\bhm\ relation is
that, under the assumption of a universal PSD shape in AGN, then \nbh$\sim
1.7\times 10^{-6}/$M$_{7}$ Hz. It is interesting to see how this relation
compares with the results from the power spectral analysis of recent
observations in AGN.  Table~1 lists the \bhm\ and \nbh\ estimates for all
AGN where a ``$-2$ to $-1$" frequency break has been detected recently,
irrespective of whether the same object has a long term \rms\ measurement
or not. The \bhm\ of NGC~5506 was estimated using the stellar velocity
dispersion measurement of 180 km/sec (Oliva \etal, 1999), together with
the \bhm\ -- stellar velocity dispersion relation of Tremaine \etal\ 
(2002). 

%%%%%%%%%%%%%% Fig. 2 %%%%%%%%%%%%%%%%%%
\begin{figure}
\psfig{figure=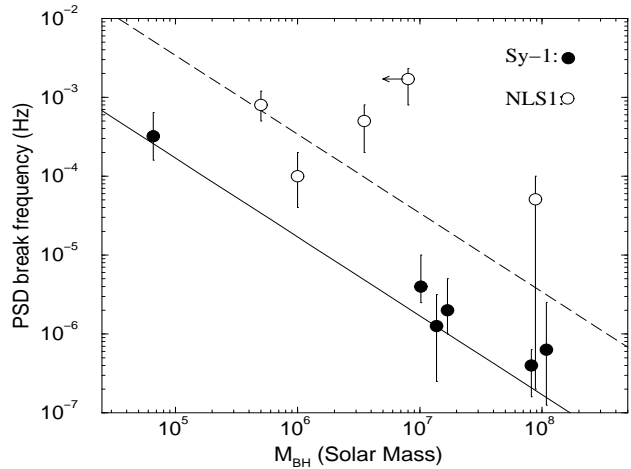,height=7.0truecm,width=9.5truecm,angle=0,%
bbllx=20pt,bblly=10pt,bburx=570pt,bbury=460pt} 
\caption[]{PSD Break frequency plotted as a function of \bhm. The solid
line shows a plot of the \nb$=C/M_{7}$ relation, with $C$ being equal to
the value from the best model fitting of the \rms--\bhm\ relation. The
dash line shows a plot of the same relation with $C$ being 20 times
larger.}
\end{figure}
%%%%%%%%%%%%%%%%%%%%%%%%%%%%%%%%%%%%%%%%

Fig.~2 shows a plot of \nbh\ as a function of \bhm. The solid line is {\it
not} a plot of the best fitting model to the data. Instead, it shows a
plot of the $\nu_{bf}=C/$M$_{7}$ relation, with $C=1.7\times 10^{-6}$ Hz
(i.e. the value estimated from the best model fitting to the \rms\ vs
\bhm\ data in Section~2). The agreement between this line and the BLS1
data is very good. However, the NLS1s \nbh\ estimates lie consistently
above the solid line. The dashed line in Fig.~2 shows a plot of the
$\nu_{bf}=C/$M$_{7}$ with $C=20\times 1.7\times 10^{-6}$ Hz (i.e. the $C$
value which can explain the high \rms\ measurement of NGC~4051, as
discussed in Section~2.2). As Fig.~2 shows, the agreement between this
line and the \nbh\ measurements of NLS1s is good. Therefore, both the
large value of the \rms\ measurement of NGC~4051 and the location of the
NLS1 data in Fig.~2 suggest that, for the same \bhm, \nbh\ is $\sim$ 20
times higher in NLS1s than in BLS1s.

%%%%%%%%%%%%%%%%%%%%%%%%%%%%%%%%%%
\subsection{A universal \psdamp\ value in AGN?}
%%%%%%%%%%%%%%%%%%%%%%%%%%%%%%%%%%

Another result from the study the \rms--\bhm\ relation is that PSD$_{amp}$
should be $\sim 0.017$ in all AGN.  It is not straightforward to compare
this result with recent results from power spectral analysis of X--ray
light curves, as in most cases the PSD normalisation is not reported. In
the cases where the relevant information is listed, the estimated \psdamp\
values are indeed close to 0.017.  For example, using the results of
Markowitz \etal\ (2003), one can find that: \psdamp=0.021, 0.016, 0.012,
0.016, and 0.017 in the case of F9, NGC~5548, NGC~3783, NGC~3516, and
NGC~4151, respectively. In the case of Cyg X-1, the Belloni \& Hasinger
(1990) data suggest that \psdamp$=0.02\pm 0.002$. This is consistent,
within the errors, with the value of 0.017. Interestingly, the existing
data for NLS1s are systematically lower than 0.017. For example, Papadakis
\etal\ (2002) and Vaughan \& Fabian (2003) find \psdamp$\sim 0.005$ and
$\sim 0.01$ in the case of Ark~564 and Mrk~766, respectively. 
Furthermore, McHrady \etal\ (2003) find \psdamp$\sim 0.01$ for NGC~4051.
However, this value was estimated using a PSD model different than the
``broken power-law model" used in the other cases. In fact, as Figure~17
in McHardy \etal (2003) shows clearly, the \psdamp\ of NGC~4051 is
probably as large as the \psdamp\ of other BLS1s.

%%%%%%%%%%%%%%%%%%%%%%%%%%%%%%%%%%%%%%%%
\subsection{The accretion rate of AGN}
%%%%%%%%%%%%%%%%%%%%%%%%%%%%%%%%%%%%%%%

%%%%%%%%%%%%%% Fig. 3 %%%%%%%%%%%%%%%%%%
\begin{figure}
\psfig{figure=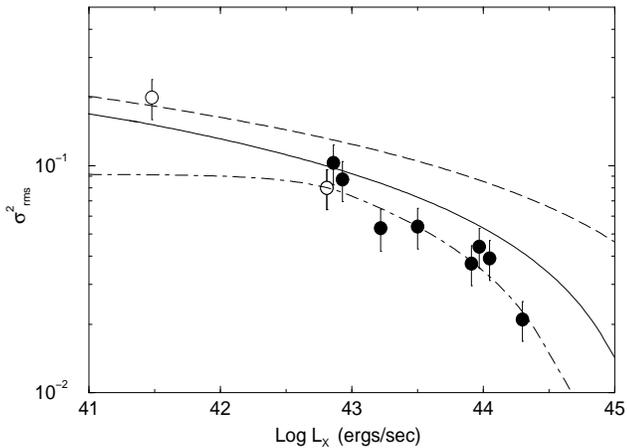,height=7.0truecm,width=9.5truecm,angle=0,%
bbllx=30pt,bblly=10pt,bburx=580pt,bbury=400pt}
\caption[]{The excess variance measurements plotted as a function of the
X--ray ($2-10$ keV) luminosity. The dot-dashed line shows a plot of the
expected \rms--$L_{X}$ relation using the results from the best model
fitting of the \rms--\bhm\ relation (shown with the dot-dashed line in
Fig.~1), and assuming that $L_{X}\propto $\bhm. The solid and dashed lines
show plots of models defined to account for the high \rms\ measurement of
NGC~4051 (see text for details).
}
\end{figure}
%%%%%%%%%%%%%%%%%%%%%%%%%%%%%%%%%%%%%%%%

It is interesting to investigate if the \rms\ vs $2-10$ keV X--ray
luminosity ($L_{X}$) correlation of Markowitz \& Edelson (2001) is in fact
a byproduct of the \rms\ vs \bhm\ correlation presented in this work. This
is possible only if there exists a relation between \bhm\ and $L_{X}$,
which is the same for all the objects. This implies that the accretion
rate, when considered as a fraction of the Eddington limit, should be
similar in all objects.

Fig.~3 shows the \rms--$L_{X}$ relation for the objects with \rms\
measurements listed in Table~1, using the $L_{X}$ measurements of
Markowitz \etal\ (2003) for F9, NGC~5548, NGC~3783, NGC~3516, and
NGC~4151, Markowitz \& Edelson (2001) for 3C120, Akn120, and MCG~-6-30-15,
McHardy \etal\ (2003) for NGC~4051, and the measurement of Papadakis
\etal\ (2003) for \pg.  The dot-dashed line in Fig.~3 shows a plot of a
model, which is based on: (1) the best fitting ``2 PSD breaks with
\nbl/\nbh=30" model to the \rms--\bhm\ relation (the one shown with the
dot-dashed line in Fig.~1), and (2)  the relation $L_{X}=({\rm BH}_{\rm
Mass}/{\rm M}_{\odot})\times 10^{35.8}$ ergs/sec. As Fig.~3 shows clearly,
this line fits the data very well ($\chi^{2}=8.8/9$ dof), except from
NGC~4051. Using the mean bolometric conversion factor of $L_{bol}\sim
27\times L_{X}$ of Padovani \& Rafanelli (1988), the adopted relation
between $L_{X}$ and \bhm\ implies that $L_{bol}=1.7\times 10^{37} ({\rm
BH}_{\rm Mass}/{\rm M}_{\odot})$ ergs/sec. Since $L_{\rm Edd}=1.25\times
10^{38} ({\rm BH}_{\rm Mass}/{\rm M}_{\odot})$ ergs/sec, the main
conclusion is that the \rms--$L_{X}$ correlation could be explained in
terms of the \rms--\bhm\ relation if all the objects in the present sample
radiate at $\sim 13.5\%$ of their Eddington luminosity.

The fact that NGC~4051 is not consistent with the dot-dashed line plotted
in Fig.~3 implies that either its accretion rate is significantly larger
than $\sim 10\%$ of the Eddington limit, or its \rms\ measurement is
significantly higher than that expected for its luminosity. The solid line
in Fig.~3 shows the expected \rms--$L_{X}$ relation, using: (1) the best
fitting ``1 PSD break with \nbh\ $\propto $ \bhm$^{-1}$" model to the
\rms--\bhm\ relation (shown with the solid line in Fig.~1) and (2)  
assuming an accretion rate three times higher than 0.1 of the Eddington
limit. The agreement between the model and NGC~4051 is reasonably good.
The dashed line in Fig.~3, shows the expected \rms--$L_{X}$ relation,
using: (1) the best fitting ``1 PSD break with \nbh\ $\propto $
\bhm$^{-1}$ and $C$ 20 times larger" model to the \rms--\bhm\ relation
(dashed line in Fig.~1) and (2)  assuming an accretion rate $\sim 0.1$ of
the Eddington limit. This model also agrees well with NGC~4051. Therefore,
the ``unusually" large variability amplitude of NGC~4051 for its
luminosity does not necessarily imply an exceptionally high accretion
rate; a PSD break at a frequency higher than what is expected for its
\bhm\ can also account for the discrepancy of NGC~4051, with respect to
the other objects in the sample.

%%%%%%%%%%%%%%%%%
\section{Discussion and conclusions}
%%%%%%%%%%%%%%%%%

The main results of this work are the following: 

\noindent
(i) There is a significant correlation between the long term excess 
variance of AGN and their \bhm, in the sense that \rms\ decreases with 
increasing \bhm.

\noindent
(ii) The \rms--\bhm\ relation is consistent with the hypothesis of a
universal PSD shape in AGN with two frequency breaks, \nbh\ and \nbl. The
PSD changes its slope from $-2$ to $-1$ at \nbh, and from $-1$ to zero at
\nbl.

\noindent
(iii) The high frequency break, \nbh, decreases with increasing \bhm\ as
\nbh$=1.7\times 10^{-6} {\rm M}_{7}^{-1}$ Hz. The low frequency break,
\nbl, is $\sim$ 10 to 30 times smaller than \nbh.

\noindent
(iv) The PSD peak amplitude in the $\nu\times P(\nu)$ space is $\sim 
0.017$ in all objects.

\noindent
(v) The correlation between \rms\ and $L_{X}$ is a byproduct of the
\rms--\bhm\ correlation, with AGN radiating at $\sim 10-15\%$ of the
Eddington luminosity.

\noindent
(vi) The \rms\ measurement of NGC~4051, a NLS1 object, is larger than what
is expected for its \bhm\ and $L_{X}$. This result can be explained if its
PSD has only one break (at frequencies higher than $\sim 4\times 10^{-8}$
Hz) which is located at a frequency $\sim 20$ times higher than the
frequency expected in the case of a BLS1 with the same \bhm. An
intrinsically higher PSD amplitude (e.g. see McHardy et al. 2003) may also
contribute to the larger $\sigma^{2}_{rms}$ of this source.

These results are based on \rms\ measurements of long light curves in
the $2-10$ keV band. Since the power spectra of individual AGN appear to
be energy dependent (e.g. Papadakis \& Lawrence 1995, Nandra \& Papadakis
2001, Vaughan \etal\ 2003), if there is indeed a ``universal" PSD shape,
this should also be energy dependent. For example, a slope of $-2$ should
characterize the AGN power spectra above the high frequency break in the
$2-10$ keV band only.

The results listed above are entirely consistent with the recent results
from the power spectral analysis of \xte, \xmm, and \asca\ light curves.
For example, in all cases where a high frequency PSD slope break has been
detected, the slope of the PSD below and above this break is roughly equal
to $-1$ and $-2$ (see e.g. Uttley \etal, 2002; Markowitz \etal, 2003).
Furthermore, as discussed in Section 2.3, the \nbh--\bhm\ relation
determined from the \rms--\bhm\ correlation is entirely consistent with
the existing data.  In fact, the relation of
$T_{break}(=1/$\nbh)=\bhm$/(10^{6.5}$ M$_{\odot})$ days, of Markowitz
\etal\ (2003), is in agreement with the results of the present work.
Finally, the observed \psdamp\ values are also consistent with the value
of 0.017, which is derived from the study of the \rms--\bhm\ relation.

One of the most interesting results presented in this work is that
the $2-10$ keV band PSD of BLS1s, at least, has a ``universal" shape:  
there two break frequencies, the slope between them is $\sim -1$, the
slope above the high frequency break $\sim -2$, the high frequency break
is mass dependent, but the \psdamp\ is not. Instead it has a constant
value of $\sim 0.02$. Since this result is based on the measurement of
\rms\ and \bhm\ of ten objects only (most of which are among the most
frequently observed AGN in X--rays) it is then premature to accept that it
holds for the Seyfert galaxies as a class. Even if it does, it is not
expected that either the high frequency slope slope for example or
\psdamp\ will be exactly equal to $-2$ and $0.02$ for each individual AGN.
Instead, these values should be considered as ``average" or ``typical"
values around which the respective values of all AGN will be distributed,
in the same way that the energy spectral slopes of AGN are distributed
around the ``canonical" value of $\Gamma\sim 1.9$. Confirmation of this
result can be achieved in two ways. First, by using the results from the
PSD analysis of many AGN. This is the direct and most powerful method, as
it will allow us to actually determine the distribution of the PSD slope
and \psdamp\ values around the ``canonical" values. However, its rather
unlikely that the number of AGN with well studied PSDs will increase
significantly in the near future.  What the present works shows clearly is
that there is a second way, which involves the \rms\ estimation of many
objects with known \bhm. Although the \rms\ of an individual object cannot
constrain its PSD shape in any way, the study of the \rms--\bhm\ relation
can be useful in order to investigate whether there is indeed a
``universal" PSD shape in AGN.  Furthermore, contrary to the former
possibility, the use of archival data of many objects for the detailed
study of the \rms--\bhm\ relation should be possible in the near future as
discussed in the Introduction.

If indeed the PSD of BLS1s AGN has a ``universal" shape (in the sense that
was described above) then this result could constrain physical models
proposed to explain the physical process responsible for the X--ray
variations.  For example, models should be able to predict the \nbh--\bhm\
relation, and explain the ``universal" \psdamp\ value of $\sim 0.02$ in
AGN (perhaps in GBHs as well). As an example of how this can be achieved,
I present below a simple ``exercise", based on the assumption that the
X-ray variations of AGN at frequencies between \nbl\ and \nbh\ are
associated with accretion disc instabilities which cause variations either
to the energy release in the innermost regions of the disc or to the soft
photons input to the X--ray emitting corona.

As a first step, one should investigate the relevance of physical
processes to the origin of X--ray variability by comparing the various
physical time-scales in accretion disc to the PSD break time-scales. Let
us consider a BLS1 with \bhm$=10^{7}$ M$_{\odot}$. According to the
\nbh--\bhm\ relation, the break time-scale for this object should be
$T_{break}=6.8^{+7.6}_{-3.8}$ days. Let us also consider the case of a
standard $\alpha-$disc (Shakura \& Sunyaev, 1973) around the central
object, and let us assume that $T_{break}$ is associated with one of the
characteristic time-scales of the disc, i.e. the orbital ($T_{orb}$),
thermal ($T_{th}$), sound-crossing ($T_{sc}$), and viscous time-scale
($T_{visc})$.  Using the relations given by Treves, Maraschi, \&
Abramowicz (1988), one can estimate these time-scales at every radius from
the central object. The results show that $T_{break}\sim T_{orb}$ at $R
\sim 70-80 R_{S}$ (where $R_{S}$ is the Schwarzschild radius). However,
this seems rather unlikely as most of the energy in the accretion disc is
dissipated at smaller radii. $T_{break}$ could also correspond to $T_{sc}$
at $R\sim 3-5R_{Sch}$ (assuming that $R\sim 100 H$, where $H$ is the
disc's scale height). However, if $T_{break}$ for a NLS1 with the same
\bhm\ is $\sim 20$ times faster, {\it and} $T_{break}$ corresponds to the
{\it same} physical time-scale in both type of objects, then obviously
this cannot be the sound-crossing time-scale. Furthermore, $T_{visc}$ is
far too slow to account for $T_{break}$. Assuming that $\alpha\le 0.1$
(where $\alpha$ is the viscosity parameter), and $R\le H$, then $T_{visc}
> 870$ days even at $R=3R_{S}$. Finally, $T_{break}$ could correspond to
$T_{th}$ at $R\sim 30 - 40 R_{Sch}$, for $\alpha=0.1$. In the case of a
NLS1 with the same \bhm, $T_{break}$ could also correspond to $T_{th}$,
but at a smaller radius ($\sim 4-6 R_{S}$).

Viscosity variations which happen at large radii, and develop on the local
$T_{visc}$, can affect the energy release of the accretion disc in its
innermost part, and result in a PSD with slope $-1$ between \nbl\ and
\nbh\ (Lyubarskii, 1997).  However, if the PSD breaks are associated with
the thermal time-scales, then perhaps as the accretion rate increases at
various radii and the disc becomes unstable, thermal instabilities set
in. The disc temperature will increase, increasing the local flux, hence
the soft photons input to the X--ray corona. These soft photon ``flares"
should then develop on the local thermal time-scale. Assuming an
exponential increase of soft photons with time, the superposition of these
flares with the different time-scales can lead to a PSD with a flat shape
up to \nbl$=1/T_{th}(R_{out})$, a $-1$ slope up to
\nbl$=1/T_{th}(R_{in})$, and $-2$ slope at higher frequencies (Lehto
1989), where $R_{out}$ and $R_{in}$ are the outer and inner radius of the
disc region which is affected by the thermal instabilities. If $R_{out}=200
- 300 R_{S}$, and $R_{in}=30-40 R_{S}$, then the resulting PSD will be in
agreement with the present results.

The fact that the $P(\nu)\times\nu$ value between \nbl\ and \nbh\ is
similar in all objects implies that the PSD integral between \nbl\ and
\nbh\ (i.e. the excess variance associated with the components of those
frequencies) remains roughly constant, irrespective of \bhm. In the
context of shot noise models where flares occur randomly with a constant
rate of say $\lambda$ flares per unit time, and have the {\it same} shape
and amplitude, then \rms$\propto 1/\lambda$. Considering the case of soft
photons flares, which develop on the local thermal time-scale, as \bhm\
increases, and $T_{th}$ increases accordingly, it is natural to expect
that $\lambda$ should decrease, as it should take longer time for the
instability to develop and the accretion disc to ``relax" before the next
instability starts to build up. In this case we should expect \psdamp\ to
increase with increasing \bhm. However, at the same time, as \bhm\
increases, the flares should become ``smoother" (because $T_{th}$ also
increases). This fact should result in a decrease in \rms. Perhaps then,
because an increase of the \bhm\ causes a decrease of $\lambda$ and, at
the same time, ``smoother" flares, and because these two effects affect
the observed variability in the opposite way, \rms\ may turn out to be
roughly constant in all objects. However, further work is needed in order
to examine if such a model can also predict correctly the \psdamp\ value
of $\sim 0.02$, as suggested by this work.

The differences in the variability properties between NLS1s and BLS1s can
be explained by the fact that \nbh\ is higher in a NLS1 than in a BLS1
with the same \bhm. The same explanation has also been proposed recently
by McHardy \etal\ (2003). In the framework of the picture presented above,
which involves thermal instabilities at a certain region of the accretion
disc, the different \nbh\ values should correspond to differences in the
location of this region in the two type of objects. In the innermost
region of BLS1s (i.e. at radii smaller $30 R_{S}$) the thermal
instabilities could be stabilized by a physical mechanism (e.g. the
formation of a hot wind). The same mechanism may not be sufficient to
stabilize the innermost region of the NLS1s, perhaps because the accretion
rate is larger in these objects. Whether NLS1s show a second PSD break at
\nbl\ $\sim 10-30$ times smaller than \nbh\ is not certain yet. The
results of this work show that the long term \rms\ measurement of NGC~4051
is not consistent with the presence of a second break, in agreement with
the results of McHardy \etal (2003). As these authors point out, NGC~4051
is the first Seyfert galaxy which shows a PSD similar to the power
spectrum of GBHs at high/soft state. However, the \rms\ measurement of
MCG~-6-30-15 is consistent with the presence of a second break. Ark 564
also shows two breaks in its PSD (Pounds \etal, 2001; Papadakis \etal,
2002). Power spectrum analysis of more NLS1s or the study of the
\rms--\bhm\ relation of a large number of NLS1s is necessary to clarify
this issue.

\section*{Acknowledgments}
                                                                                
The author would like to thank the referee, P. Uttley, for useful 
suggestions and comments.

\end{document}